*In situ* He$^+$ irradiation of the double solid solution (Ti$_{0.5}$,Zr$_{0.5}$)$_2$(Al$_{0.5}$,Sn$_{0.5}$)C MAX phase: defect evolution in the 350-800 °C temperature range

B. Tunca[1,2*], G. Greaves[3], J.A. Hinks[3], P.O.Å. Persson[4], J. Vleugels[2], K. Lambrinou[1,3]

[1] SCK CEN, Boeretang 200, B-2400 Mol, Belgium

[2] Department of Materials Engineering, KU Leuven, Kasteelpark Arenberg 44, B-3001 Leuven, Belgium

[3] School of Computing and Engineering, University of Huddersfield, Huddersfield HD1 3DH, UK

[4] Thin Film Physics, Department of Physics, Chemistry and Biology (IFM), Linköping University, SE-581 83 Linkoping, Sweden

*Corresponding author: B. Tunca

E-mail address: bensu.tunca@kuleuven.be

## Abstract

Thin foils of the double solid solution (Zr$_{0.5}$,Ti$_{0.5}$)$_2$(Al$_{0.5}$,Sn$_{0.5}$)C MAX phase were *in situ* irradiated in a transmission electron microscope (TEM) up to a fluence of 1.3×10$^{17}$ ions·cm$^{-2}$ (~7.5 dpa), using 6 keV He$^+$ ions. Irradiations were performed in the 350-800 °C temperature range. *In situ* and post-irradiation examination (PIE) by TEM was used to study the evolution of irradiation-induced defects as function of dose and temperature. Spherical He bubbles and string-like arrangements thereof, He platelets, and dislocation loops were observed. Dislocation loop segments were found to lie in non-basal-planes. At irradiation temperatures ≥ 450 °C, grain boundary tearing was observed locally due to He bubble segregation. However, the tears did not result in transgranular crack propagation. The intensity of specific spots in the selected area electron diffraction patterns weakened upon irradiation at 450 and 500 °C, indicating an increased crystal symmetry. Above 700 °C this was not observed, indicating damage recovery at the high end of the investigated temperature range. High-resolution scanning TEM imaging performed during the PIE of foils previously irradiated at 700 °C showed that the chemical ordering and nanolamination of the MAX phase were preserved after 7.5 dpa He$^+$ irradiation. The size distributions of the He platelets and spherical bubbles were evaluated as function of temperature and dose.

## Introduction

The MAX phases are nanolaminated ceramics with the $M_{n+1}AX_n$ general stoichiometry, where typically '$M$' corresponds to an early transition metal, '$A$' is an element from groups 13-15 in the periodic table, '$X$' is either C or N, and n = 1, 2 or 3 [1]. For nuclear applications close to the reactor core, N is not preferred as an $X$ candidate element to avoid formation of the long-lived isotope $^{14}$C. The hexagonal close-packed (hcp) unit cell of the MAX phases consists of '$n$' number of $M_6X$ octahedral "ceramic" building blocks interleaved by single atomic layers of the "metallic" '$A$' element. Solid solutions on the $M$ and $A$ sites broaden the compositional range of the MAX phases that can be synthesised and tailored in order to meet the property requirements of the targeted application. Moreover, the formation of double solid solutions (both on the $M$ and $A$ sites) allows the experimental synthesis of nearly phase-pure MAX phase ceramics, by adjusting the relative size of the $M$ and $A$ atoms and thus modifying the crystal lattice distortions [2,3]. Ceramics based on the double solid solution $(Zr_{0.8},Nb_{0.2})_2(Al_{0.5},Sn_{0.5})C$ [2] and $(Zr_{0.5},Ti_{0.5})_2(Al_{0.5},Sn_{0.5})C$ [3] MAX phases are characterised by improved phase purity over that of the 'starting' $Zr_2AlC$ ceramic (highest achieved purity: 67 wt.% MAX phase, 33 wt.% $ZrC_x$ [4]); in particular, the $(Zr_{0.5},Ti_{0.5})_2(Al_{0.5},Sn_{0.5})C$ MAX phase ceramic comprises 98 wt.% MAX phase and 2 wt.% intermetallic compounds (IMCs). Phase purity is regarded as a key design criterion for nuclear fuel cladding materials, as phase mixtures are prone to in-service disintegration due to differential radiation swelling and/or differential thermal expansion that may lead to microcracking.

For the development of advanced nuclear materials, such as accident-tolerant fuel claddings, the assessment of radiation tolerance is of paramount importance. In this respect, He irradiation studies have been proven a very useful tool, especially with respect to the investigation of the swelling behaviour of innovative nuclear materials. High-energy $\alpha$-particles produced by (n, $\alpha$) decay can result in He bubble formation, microstructural changes, void swelling, blistering, embrittlement, and surface roughening in irradiated materials; these effects are mainly caused by the low solubility and high mobility of He atoms in metals and ceramics [5,6]. Unlike other irradiation defects, such as interstitials and vacancies that can recombine or annihilate, He atoms must be accommodated/stored in the material in a stable form, which should not grow into large voids that will unavoidably degrade material properties [7]. The effects of He ion irradiation on other ceramics, such as SiC, $Si_3N_4$ and $Al_2O_3$, have been studied, and the formation of cavities, cavity clusters situated on dislocation loops, and two-dimensional platelet-like formations have been typically observed [8]. According to existing literature, MAX phase ceramics are no exception, and the formation of He bubbles of various shapes has been reported [9]. Understanding how He is incorporated in the MAX phases is, therefore, crucial for the optimization of MAX phase-based ceramics for specific nuclear applications. Furthermore, ion irradiation is a high throughput, fast and efficient material screening tool that allows the attainment of high dpa (displacements per atom) levels within reasonable times and at a much lower cost compared to neutron irradiations. Helium ion irradiations can, thus, give valuable insights on the response of specific candidate materials to in-reactor He generation.

The MAX phases are steadily gaining attention for select envisaged nuclear applications, and a number of He irradiation studies on the MAX phases have already been reported in the open literature – for example on $Ti_3AlC_2$ [5,9–16], $Ti_2AlC$ [17–19], $Ti_3SiC_2$ [6,20–22], $Cr_2AlC$ [23,24], $V_2AlC$ [25], and $Ti_4AlN_3$ [19]. Overall, the He irradiation of MAX phase compounds tends to result in He atoms residing preferentially in the *A*-layer, thereby dislocating *A*-elements from their original sites [6,10,12,13,18,20,21]. First-principles calculations for both $Ti_3SiC_2$ and $Ti_3AlC_2$ suggested that the *A*-layers are the preferred sites for He atoms to reside [6,10,13,17].

First-principles calculations have been used to study various shapes of He clusters reported in literature as ranging from spherical bubbles to platelets [13]. In $Ti_3AlC_2$, the He atom trapped in an Al vacancy is expected to promote secondary Al vacancy formation that leads to the trapping of more He atoms along the Al layers. Such confined growth of He complexes leads to the formation of He platelets. In agreement with this finding, the energy barrier for He atom migration along the MAX phase basal planes was reported to be much smaller (0.294 eV) than that for out-of-plane migration and parallel to <0001> (2.980 eV). While a single Al vacancy was reported to accommodate up to ten He atoms and induce secondary Al vacancy formation allowing platelet growth, a C vacancy can accommodate three He atoms and promote the 3D growth of He clusters [13]. Helium atoms on C vacancies, contrary to Al vacancies, do not promote further C vacancy formation, which suggests that He atoms will not cluster along C-layers, but will have to expand to other layers for growth, leading to spherical bubbles [13]. The solution energy of He in Al vacancies (0.944 eV) is smaller than that in C vacancies (2.035 eV), which makes the trapping of He in Al layers more likely to occur [13]. This is, of course, directly influenced by the initial vacancy concentration in the pristine (non-irradiated) $Ti_3AlC_2$.

According to calculations, He atoms will rapidly diffuse into the Si layers in $Ti_3SiC_2$ above 500 °C [20]. Compared to many metals, where at least five He atoms are required to displace a host atom forming a lattice vacancy [26–28], one He atom alone can displace a Si atom in $Ti_3SiC_2$, while Ti and C are more stable [29]. This will allow anisotropic He cluster/platelet growth in the Si layers. The loss of Si from its original lattice site may lead to degradation of the MAX phase with TiC formation and confined He platelets that can accelerate cleavage fracture. Upon He desorption during $Ti_3SiC_2$ annealing at high temperatures, Si interstitials have been predicted to reoccupy vacated sites [6].

Experimental findings have already indicated overall trends in the He irradiation response of the MAX phases. At room temperature (RT), the formation of spherical He bubbles has been reported [9,11,12,18,21,25,30]. In some studies, linear arrangements of spherical He bubbles seen in projection have been observed [9,30]; such defects are hereafter referred to as "string-like" He bubbles, for brevity and consistency with the existing literature. Antisite defects, especially between *A* and *M* atoms, form easily and may lead to the disruption of chemical ordering, i.e., the loss of the nanolaminated MAX phase crystal structure [11,12,21,23,24,30]. This disruption is commonly accompanied by the formation of TiC in $Ti_3AlC_2$ and $Ti_3SiC_2$ [5,9,21] and a $Cr_2C$-like structure in $Cr_2AlC$ [23]. However, a recent study [31] showed the formation of *fcc*-$(M_{n+1},A)X_n$ carbides as result of MAX phase (neutron and ion) irradiation, instead of the erroneously perceived formation of binary carbides *(MX)* previously reported in literature [32–35]. This irradiation-induced phase transformation from *hcp*-$M_{n+1}AX_n$ to *fcc*-$(M_{n+1}A)X_n$ has been confirmed with He and Au irradiations performed on $Ti_3AlC_2$, $Ti_2AlC$, $Ti_3SiC_2$, $Nb_4AlC_3$, $V_2AlC$, $Ti_4AlN_3$, and $Ti_2AlN$ MAX phases [31]. Cracks may form during RT irradiations due to lattice distortions, i.e., a contraction along the <*a*> axis and an expansion along the <*c*> axis, which are minimized during irradiation at elevated temperatures [9,15,18,30]. Cracks were reported to nucleate at string-like and plate-like He bubbles [9,30]. Upon irradiation above 300 °C, the He bubbles tend to acquire plate-like shapes [9,18,21]. Irradiations at higher temperatures result in less damage, due to the dynamic recovery of defects. No amorphisation has ever been reported during He ion irradiation of MAX phase compounds in the temperature range from RT to 750°C [17] and doses up to 52 dpa [9].

The work reported here aims at investigating the He irradiation tolerance of the quasi phase-pure double solid solution $(Zr_{0.5},Ti_{0.5})_2(Al_{0.5},Sn_{0.5})C$ MAX phase ceramic at reactor-relevant temperatures for both Gen-II/III light water reactors (LWRs; nominal operation conditions and high-temperature transients) and Gen-IV lead-cooled fast reactors (LFRs; nominal operation conditions). The evolution of defects and microstructural changes were monitored by *in situ* TEM as functions of damage dose and temperature. The experimental findings improve the collective knowledge on MAX phase radiation tolerance and can be used as a guide to the further optimization of advanced nuclear materials based on the MAX phases, especially fuel cladding materials and/or oxidation/corrosion-resistant coatings thereof.

## Experimental

### Synthesis and characterization

The experimental synthesis and microstructural characterization of the double solid solution $(Zr_{0.5}Ti_{0.5})_2(Al_{0.5},Sn_{0.5})C$ *(ZTAS)* MAX phase-based ceramic ion-irradiated in this work has been reported previously [3]. In brief, this ceramic was produced by reactive hot pressing of $ZrH_2$, $TiH_2$, Al, Sn and C powders at 1450 °C for 30 min under 30 MPa, using a starting powder stoichiometry of $M_2A_{1.1}C_{0.95}$ and Zr:Ti and Al:Sn ratios of 50:50. Phase identification was performed by X-ray diffraction (XRD) using Cu Kα radiation at 30 kV and 10 mA in a Bragg–Brentano geometry (Bruker D2 Phaser, BRUKER) in the 5–75° 2θ range with a step size of 0.02° and a time of 0.2 s per step. Electron backscattered diffraction (EBSD) measurements were performed with a Nova NanoLab 600 DualBeam (FEI, FIB/SEM) equipped with an EBSD detector. Characterization of the pristine material can be found in the supplementary information.

Thin foils for TEM examinations during *in situ* ion irradiation experiments and post-irradiation examination (PIE) analyses were prepared by focused ion beam (FIB; Nova NanoLab 600 DualBeam, FEI, The Netherlands) [36]. A protective Pt layer was deposited during thin foil preparation in the FIB, and ion milling was performed using 30 keV Ga ions, while the voltage was progressively decreased during thinning steps down to 5 keV for final cleaning.

### *In situ* He$^+$ ion irradiation

At the outset of the *in situ* ion irradiation experiments, the FIB foils were 'mapped' in the TEM at low magnification to find areas of interest, and selected area electron diffraction (SAED) patterns were collected from these areas prior to ion irradiation and/or heating. To simultaneously follow up the irradiation response of the *ZTAS* MAX phase and the occasionally present IMCs, areas containing both phases were selected. This enabled assessment of whether or not the IMCs were detrimental for the overall integrity of the thin foils. A heating rate of 100 °C/min was applied to reach the targeted irradiation temperatures. To assess the effect of heating/cooling on the material structural integrity, considering the anisotropic thermal expansion of the *ZTAS* MAX phase [3], a reference foil was included in the experiments performed at 350 °C and 500 °C. The reference foils were shielded from ion irradiation, but shared the same thermal history with the irradiated foils. Once the irradiation temperature was reached, SAED patterns were collected from individual grains. *In situ* ion irradiation experiments were performed in the MIAMI-2 facility [37], using a TEM (Hitachi H-9500 ETEM, Hitachi High-Technologies, Japan) operated at 300 kV. The FIB foils were irradiated using 6 keV He$^+$ ions, with an ion flux of $8.8\times10^{13}$ ions·cm$^{-2}$·s$^{-1}$. The use of such low-energy ions was necessary to implant the He ions in the thin foil, thus allowing the study of radiation swelling effects, such as the formation of phase-specific He bubbles, grain boundary crack nucleation and propagation due to differential swelling, etc. The targeted ion fluence was reached in seven consecutive steps, using one FIB foil per temperature. The dose levels in each step are given in **Table 1.** The fluence-to-dpa calculations were performed using *SRIM 2013* and the procedure proposed by Stoller *et al.* [38]. Calculations were done in the 'Ion Distribution and Quick Calculation' mode with lattice and sputter energies set to zero, while the displacement threshold energies used for the target elements were 40 (Zr), 30 (Ti), 25 (Al), 40 (Sn) and 25 (C) eV [39]. The damage profile was obtained using the recoil/damage calculations made with the Kinchin-Pease model, and the resultant total phonon distribution was converted to displacements per atom using the following equation [40,41]:

$$dpa(x) = \frac{0.4\, F_D(x)\, \Phi}{N E_d} \qquad (1)$$

where *x* is the depth from the irradiated surface, $F_D$ the deposited energy distribution, $E_d$ the average displacement energy for the *ZTAS* MAX phase, *N* the atomic density, and $\Phi$ the ion fluence. At the end of each experiment, the thin foils had a total dose of 7.5 dpa averaged across 50 nm-thick foils (**Fig. 1**). The thickness of the examined thin foil areas ranged between 55 nm and 185 nm. Measuring the TEM foil thickness was done using two different techniques: Monte Carlo simulations in combination with backscattered electron beam image analysis in the scanning electron microscope (SEM) [42], and convergent beam electron diffraction in the TEM [43]. The employed methodology (**Fig. S1**) and obtained colour-coded thickness maps (**Fig. S2**) are provided in the supplementary information.

**Table 1.** Temperature, fluence and approximate damage dose for the *in situ* He⁺ irradiations.

| Irradiation temperatures (°C) | 350, 450, 500, 550, 600, 700, 800 | |
|---|---|---|
| **Irradiation step** | **Fluence (ions·cm⁻²)** | **Damage dose (~dpa)** |
| 1 | $8.50\times10^{15}$ | 0.5 |
| 2 | $1.70\times10^{16}$ | 1.0 |
| 3 | $3.40\times10^{16}$ | 2.0 |
| 4 | $5.10\times10^{16}$ | 3.0 |
| 5 | $6.80\times10^{16}$ | 4.0 |
| 6 | $8.50\times10^{16}$ / $1.04\times10^{17}$ | 5.0 / 6.1 |
| 7 | $1.30\times10^{17}$ | 7.5 |

With the exception of the video recording between 5.0 and 5.9 dpa during the irradiation experiment at 500 °C, all ion irradiations were done with the electron beam off to avoid the interaction of effects induced by ion irradiation with effects due to electron beam irradiation/annealing. After each irradiation step, electron beam imaging was performed (at the irradiation temperature) to collect SAED patterns and bright field (BF) images at various defocus levels from the areas of interest. No tilting was performed during image collection. Slight changes in grain orientations were induced by the distortion (bending) of the thin foils as result of heating and/or irradiation-induced swelling. After data collection at the target damage dose (step-7), the thin foil was cooled to RT at a cooling rate of ~100 °C/min. Due to ion beam instability during the irradiation at 550 °C, that experiment was terminated at an unknown, but significantly higher than the targeted 7.5 dpa damage level. Due to the uncertainty in the achieved dose level, this foil was not studied in detail; however, it is described in the text as an example of radiation-induced damage at a higher than the targeted irradiation dose of 7.5 dpa.

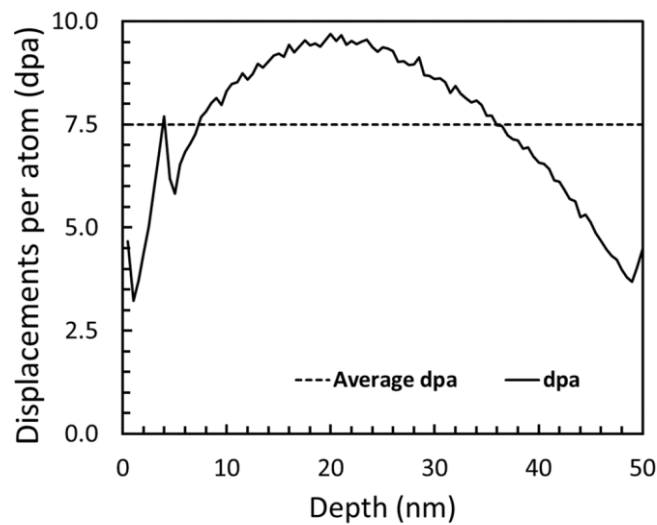

**Fig. 1.** Displacements-per-atom depth for TEM foils with an average damage dose of 7.5 dpa (indicated by the dashed line).

Post-irradiation examination (PIE)

PIE was performed on both irradiated and reference (only heat treated) thin foils to study in detail the defects generated due to irradiation and/or heating. The PIE was done on a TEM (CM200 FEG, FEI/Philips, The Netherlands) operated at 200 kV, using SAED as well as BF and dark field (DF) imaging techniques. High-resolution TEM PIE was performed for the thin foil irradiated at 700 °C, using high-angle annular dark-field scanning transmission electron microscopy (HAADF-STEM) imaging, combined with energy dispersive X-ray spectroscopy (EDX) elemental mapping on a double Cs-corrected FEI Titan[3] 60–300 operated at 300 kV and equipped with a Super-X EDX system.

For the determination of He bubble sizes, underfocused (-2 μm) BF TEM images were used, where size measurements were taken using the inside edge of the dark Fresnel fringes [44]. Approximately 200 bubbles per frame (3–5 frames for each dpa step) and all clearly visible He platelets (approximately 100-530 platelets per dpa step from multiple frames) were measured to generate a statistically relevant defect size distribution. The possibility that not all He platelets measured from the *in situ* captured images were perfectly 'edge on' might account for a small error in the determined platelet sizes. When comparing directly to bulk irradiations, it is important to note that TEM results acquired during *in situ* ion irradiation of thin foils or upon the ensuing PIE may vary due to the proximity of surfaces that can act as defects sinks.

## Results and Discussion

### Pristine material

The reaction-hot-pressed ceramic comprised 98±1 wt.% of the *ZTAS* MAX phase and minor amounts (2±1 wt.%) of the $Al_2Zr$ IMC with trace amounts of dissolved Sn and/or Ti, according to the Rietveld refinement of its X-ray diffraction (XRD) pattern (**Fig. S3**). The nanolaminated structure of the MAX phase was visible on fracture surfaces, while the electron backscatter diffraction (EBSD) orientation map indicated a slight basal texture caused during hot pressing (**Fig. S3**). The prepared FIB thin foils sometimes exhibited localized ion milling damage. As each thin foil was examined at RT prior to the $He^+$ ion irradiation, TEM images of any pre-existing damage were captured for comparison with the final (post-irradiation) defect microstructure. Whenever possible, however, grains with existing ion milling damage in the pristine (i.e., non-irradiated) condition were excluded from the areas selected for *in situ* follow-up (i.e., for data collection at each dpa step) of radiation-induced damage. An example of FIB-induced damage is shown in **Fig. S4**, where the diffuse ring pattern in the SAED pattern indicates amorphisation of an initially crystalline grain. Dislocations and occasional stacking faults (SFs) stemming from processing were present in the pristine material (**Fig. S4**). After heating and prior to irradiation, contamination appeared on both surfaces of the thin foil, usually starting at around 350 °C. SAED patterns collected from grains with such contamination did not reveal extra spots that might be attributed to another crystalline phase. The source of this contamination is likely to be hydrocarbons originating from transportation/handling of the thin foils or from the TEM itself.

He⁺ irradiation at 350 °C

**Figures 2a-2c** show select areas of interest from thin foils irradiated to 2.0, 4.0 and 7.5 dpa at 350 °C. At damage levels less than 7.5 dpa, no significant overall change in the microstructure was observed, and the grain boundaries (GBs) appeared intact. At higher magnifications, however, small He bubbles were visible, appearing as white round features with dark rims in underfocus images and dark features with bright rims in overfocus images. In light of similar experimental findings on irradiated MAX phases [9,18], these features were identified to be He bubbles.

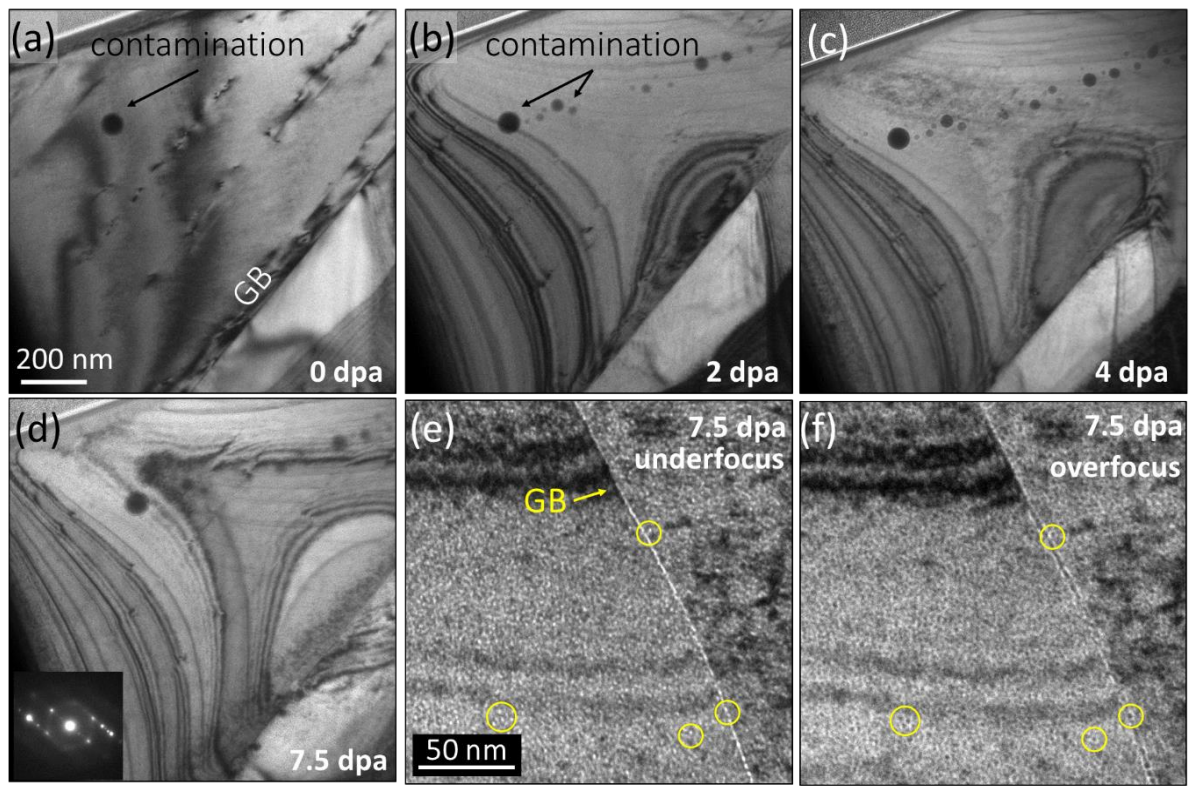

**Fig. 2.** BF TEM images of *ZTAS* grains at 350 °C (a) before irradiation, after (b) 2.0, (c) 4.0, and (d-f) 7.5 dpa. The scale bar in (a) applies to (a-d). The images in (e) and (f) show an area near a GB imaged in under- and overfocus, respectively. The circles in (e) and (f) indicate some of the He bubbles.

The term "bubble" is herein used to describe spherical features containing He, "void" is considered as a cluster of vacancies and thus implies an empty volume, and "cavity" is a collective term used for both defects [45]. Helium atoms are known, in general, to stabilize cavities, binding strongly with vacancies and attracting more of the implanted He atoms and vacancies to agglomerate and grow into bubbles [46]. In the *ZTAS* MAX phase compound, bubbles appeared uniformly distributed in the microstructure, i.e., both inside grains and at GBs, with a slight affinity for the GBs (**Figs. 2e-2f**). Bubbles were also found near GBs, indicating no bubble-denuded zones at this temperature. At the maximum dose of 7.5 dpa, GBs contained small bubbles (**Figs. 2e-2f**) that did not lead to GB tearing. Black spot defects were already observed at the smallest dose of 0.5 dpa. Due to their small size, the nature of these black spot defects (also called 'black dots') is not clearly defined in literature [46]. They are claimed to be either small Frank loops of the interstitial type [47] or vacancy/interstitial clusters [46,48,49]. The heat treated reference foil in this experiment did not show any GB tearing or microcracking.

## He⁺ irradiation at 450 °C

In the 450 °C irradiation experiment, He bubbles were already visible at the lowest dose of 0.5 dpa, and more apparent after 2 dpa, particularly at the IMC/MAX phase GBs (black ellipses in **Figs. 3a'-3b'**). **Figures 3a'-3d'** are all from the same location, marked by yellow squares in **Figs. 3a-3d**. The reduction in visibility of the edge-on GB in **Figs. 3c'-3d'** (GB marked by the black ellipses in **Figs. 3a'-3b'**) is believed to have resulted from the slight reorientation of the two adjacent grains (1 IMC, 1 *ZTAS* MAX phase) with increased damage. The reorientation of the IMC/MAX phase GB may be attributed to anisotropic swelling, which is readily appreciated by the different density and arrangement of He bubbles in the neighbouring grains. In certain locations, increasing the dose resulted in further accumulation of bubbles at the GBs (**Figs. 3e-3f**) – a process that eventually led to GB tearing (**Fig. 3f**). Tearing of GBs was first observed around 5.0 dpa. From 2.0 dpa onwards, "string-like" He bubbles, i.e., linear-like arrangements of nano-sized spherical bubbles, became visible. The yellow ellipses in the magnified sections shown in **Figs. 3a'-3d'** indicate that He bubbles coalesce into He 'strings' with increasing dose. At the maximum dose of 7.5 dpa, He platelets were also observed (**Fig. 3g**). Helium platelets have been observed in some metals (e.g., Ti, Mo, Ni), covalent materials (e.g., Si, $B_4C$, SiC, $Al_2O_3$, MgO), tritides (e.g., PdT and $ErT_2$ [50,51]), as well as the $Ti_2AlC$, $Ti_3AlC_2$ and $Ti_3SiC_2$ MAX phases [9,18,21]. For Si, $ErT_2$ and SiC, the formation and growth of these He platelets has been studied in more detail and will be touched upon in the discussion on the overall findings of this work.

Helium platelets were found to be oriented parallel to the basal planes (0001) of the MAX grains, as shown in **Fig. 3g**. Similar to the spherical He bubbles, they were identified as dark features with bright rims in overfocus and bright features with dark rims in underfocus BF images (**Figs. 4a-4b**). A 430 nm-long delamination encountered inside one of the MAX grains is believed to have formed on a pre-existing SF that acted as a preferred site for He bubble nucleation (**Fig. 4c**). According to the available literature, He platelets generate dislocation dipoles during their growth and increase in thickness by dipole expansion, which is similar to dislocation loop punching of spherical bubbles [52–55]. The accumulated He atoms between atomic layers push out the confining *MX* planes, resulting in dislocation dipole formations around each He platelet. When the adjacent planes are displaced by $d/2$, with $d$ being the interplanar spacing of the planes the He platelet lies on, dislocations form. Once the He platelet thickness '$s$' reaches $3 \times d$, the generated dislocations may escape from the He platelet. A schematic representation of the dislocation dipole expansion, reproduced from the work of Cowgill *et al.* [54], is provided in the supplementary information (**Fig. S5**).

Inspired by the work of Cowgill *et al.* [54] and also that of others [52,55], a HRTEM image of a *ZTAS* grain irradiated at 450 °C to 7.5 dpa and containing He platelets (**Fig. 4d**) has been filtered using the (0002) basal plane spot in the fast Fourier transformed (FFT) image. This filtered image (**Fig. 4e**) shows the lattice disturbances: some dislocation loops and dipoles surrounding the He platelets are visible. Since the images shown in **Figs. 4d** and **4e** were collected after ion irradiation, it is impossible to verify whether dislocation loops were present in that area prior to irradiation or not. However, it must be noted that these images resemble strongly the images one might expect from the previously proposed dipole expansion mechanisms for He platelet growth [52,54,55].

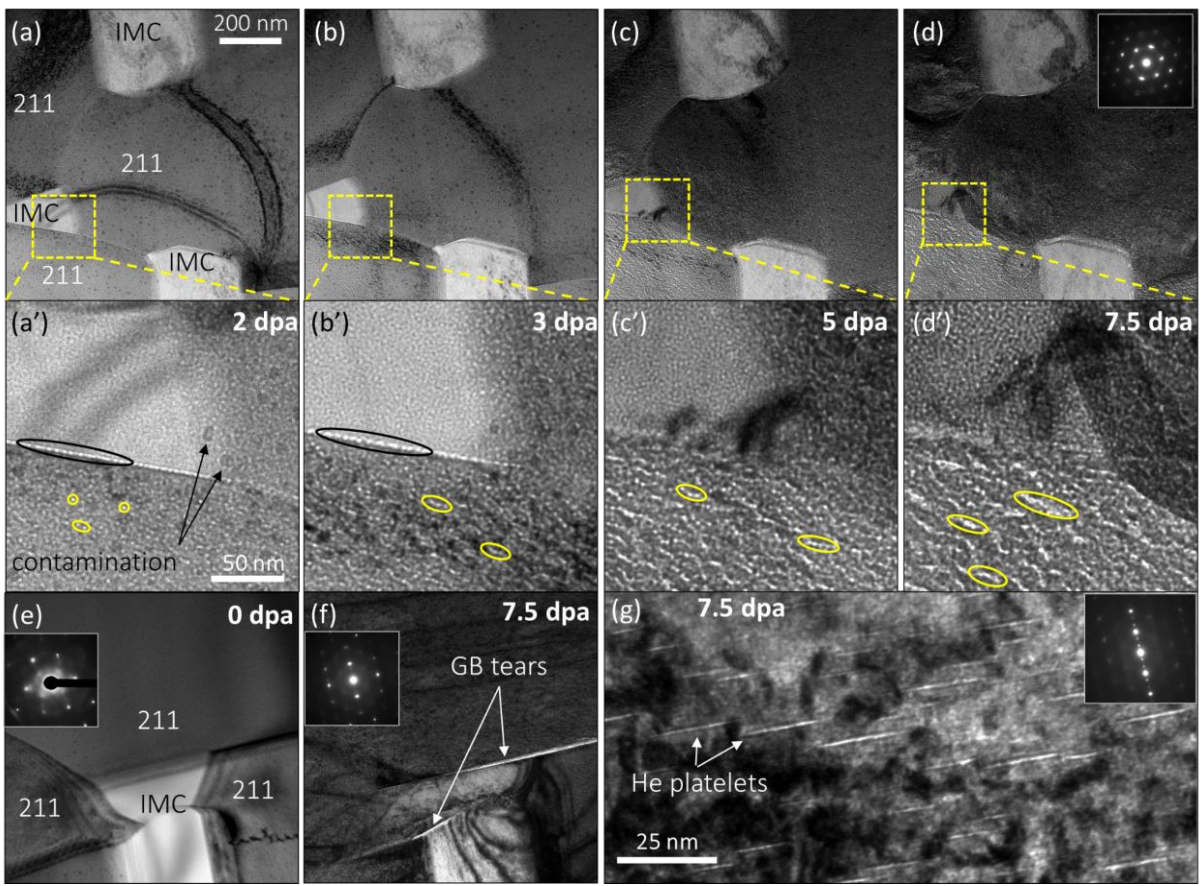

**Fig. 3.** Underfocused (-2 μm) BF TEM images collected at 450 °C at (a) 2.0, (b) 3.0, (c) 5.0, (d) 7.5, (e) 0 and (f) 7.5 dpa, with the latter showing GB tears. The scale bar in (a) applies to (a–f). Magnified images (a′–d′) of the yellow square frames in (a–d), respectively; the scale bar in (a′) applies to (a′–d′). Black ellipses in (a′-b′) indicate voids at the GBs and yellow ellipses mark alignment of bubbles into string-like formations. BF TEM image (underfocused -200 nm) in (g) shows He platelets at 7.5 dpa.

The temperature-dependent mobility of these dislocations was suggested as a critical parameter affecting the growth of He platelets. The MAX phases are characterized by basal plane dislocations that can multiply and are mobile at RT [1,56,57]. In addition, they show a brittle-to-plastic transition at high temperatures (> 900 °C [1,58]), which has been explained by the involvement of kink bands, mobile dislocation walls, GB softening or decohesion [1], underlining their temperature-dependent plasticity. In **Fig. 4e**, some of the dislocations found in the filtered image are not in the immediate vicinity of the He platelet, suggesting that dipole expansion had already occurred to a certain extent, in agreement with literature [1,56–58]. Once the dislocation loops move due to the stress built up by the implanted ions, it is likely that the strain around the He platelet is reduced. However, one should note that for the dislocation dipoles to expand, a Burgers vector with <*c*> component (less likely), dislocation climb, or cross-slip, might be required. Evidence of dislocation cross-slip from basal planes to prismatic and pyramidal planes at 900 °C has already been reported [59]. The ease of cross-slip was explained by the <*a*> type Burgers vector of the new dislocation segments that lie on non-basal planes. Dislocations, similar to GBs, act as sinks for irradiation-induced point defects, such as interstitials, vacancies, and He atoms. The interactions of these point defects with each other and with the pre-existing microstructure affect dislocation mobility [60]. The mechanism explaining the observed dislocation mobility around He platelets (**Fig. 4e**) is still not fully understood; however, it can be associated with the expected increase in dislocation mobility at 450 °C and the radiation-induced point defects. In the foil irradiated at 500 °C, mobile dislocation loops have been observed, as discussed in more detail later in the text.

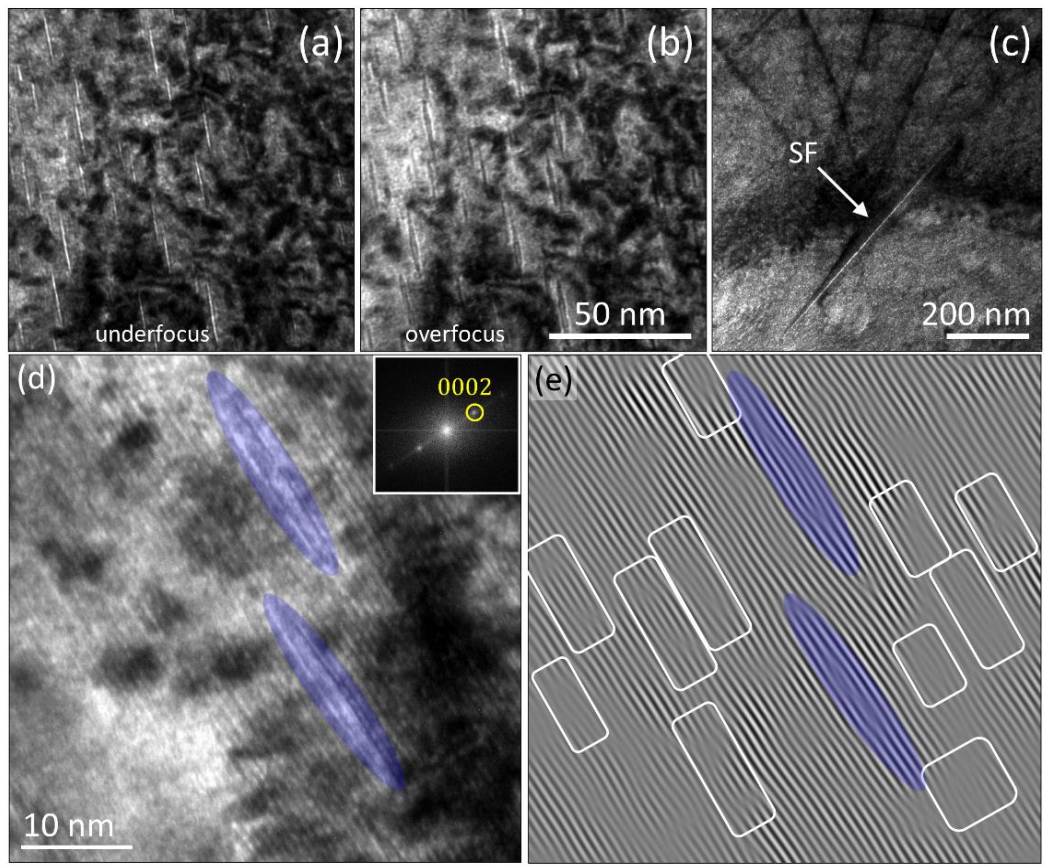

**Fig. 4.** (a) Underfocused and (b) overfocused BF TEM images of He platelets after irradiation to 7.5 dpa at 450 °C. (c) A SF acting as preferential He bubble nucleation site. (d) An area with He platelets (blue highlights), and (e) the corresponding filtered TEM image showing dislocation loops and dipoles (outlined by rectangular frames) near the He platelets.

He⁺ irradiation at 500 °C

The thin foil irradiated at 500 °C contained FIB milling damage in its pristine state, as indicated by the presence of a diffuse ring pattern, characteristic of amorphous matter, in several SAED patterns (**Fig. S4**). **Figure 5** summarises the microstructural findings observed at this temperature as a function of damage (dpa) level. Upon reaching the target temperature, contamination was observed, as indicated in **Fig. 5a**. Black spot defects and bubbles were visible starting at the smallest damage doses of 0.5–1.0 dpa. With the accumulation of bubbles at GBs, GB tearing started at around 3.0 dpa (**Fig. 5b**). At 5.0 and 7.5 dpa, string-like He bubbles, spherical bubbles, He platelets and dislocation loops were observed (**Figs. 5c-5d**). The magnified images in **Figs. 5a'-5d'** show the evolution of He bubbles with increasing dose at 500 °C, indicating no significant change in the size of spherical bubbles with dose. In the heat-treated reference foil, small cavities were observed close to the protective Pt strip, potentially due to ion milling damage and subsequent annealing. Otherwise, this non-irradiated thin foil did not demonstrate any irradiation-induced defects and all GBs remained intact.

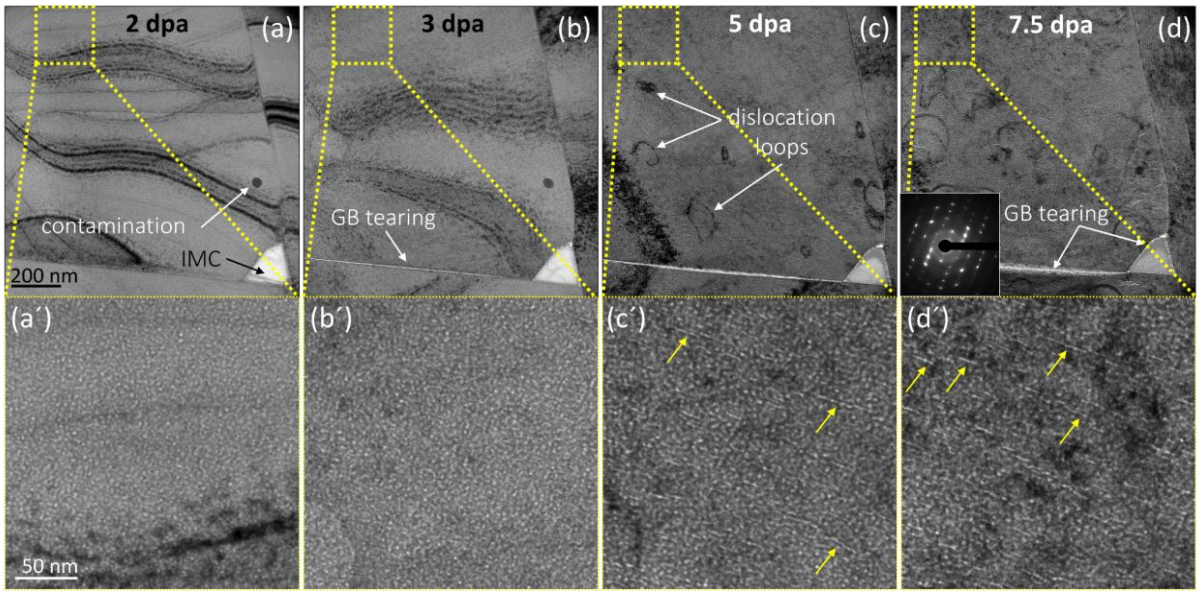

**Fig. 5.** Underfocused (-2 µm) BF TEM images collected at 500 °C at (a) 2.0, (b) 3.0, (c) 5.0 and (d) 7.5 dpa with (d) also including the corresponding SAED pattern. Magnified images of the areas in the yellow frames in (a-d) are shown in (a′-d′). The scale bars in (a) and (a′) apply to all images in the same row. The yellow arrows indicate string-like He bubbles and He platelets.

During the *in situ* examination of the area shown in **Fig. 5**, dislocation loops were observed, and some of them were identified as being mobile during electron beam imaging at a zone axis close to $[2\overline{1}\overline{1}0]$. To capture their behaviour under He ion irradiation during the dose accumulation up to ~5.9 dpa (~1.01×10$^{17}$ ions·cm$^{-2}$), a BF TEM imaging video was recorded (a 10-times accelerated version of that video is provided in the supplementary information). This video captured the nucleation, growth and disappearance of dislocation loops under He$^+$ ion irradiation. Still images from the video at different times are shown in **Fig. 6** together with the corresponding doses. After video recording, the electron beam was turned off to continue the damage dose accumulation up to 7.5 dpa. White arrows in **Figs. 6a-6d** mark the formation of new dislocation loops, while the black arrow in **Fig. 6c** points to a dislocation dipole (more clearly seen in the video in the supplementary information). The shaded area in **Fig. 6c** marks a loop that has been growing between **Figs. 6a-6b**, and the white arrow a new, smaller loop giving the impression of having formed within the larger, existing loops in projection view (**Fig. 6c**). In **Figs. 6f-6g**, typical shapes of pinned dislocations are visible. Considering the orientation of the grain, which is an off-zone axis close to the $[2\overline{1}\overline{1}0]$ zone (SAED inset in **Fig. 5d**), these loops appear to have segments lying on non-basal planes, as indicated by the inverted DF image that was obtained using the $<0\overline{1}13>$ reflection and shown in **Fig. 6h**. Non-basal dislocations, not observed during RT deformations, have been observed during deformation of Ti$_2$AlC at 900 °C [59] and in creep experiments of Ti$_3$AlC$_2$ at 900 °C [61], as well as in as-fabricated Ti$_4$AlN$_3$ [62]. It is, thus, not unlikely that He$^+$ irradiation at 500 °C generated favourable conditions for non-basal dislocation formation and motion. When examined at higher magnification, some of the loops seemed to be pinned by He platelets (**Figs. 6f-6g**). The video also reveals that there is a stepwise growth of loops, presumably due to the fact that they experienced different degrees of pinning by the existing defects, such as He bubbles, He platelets, black dots, etc.

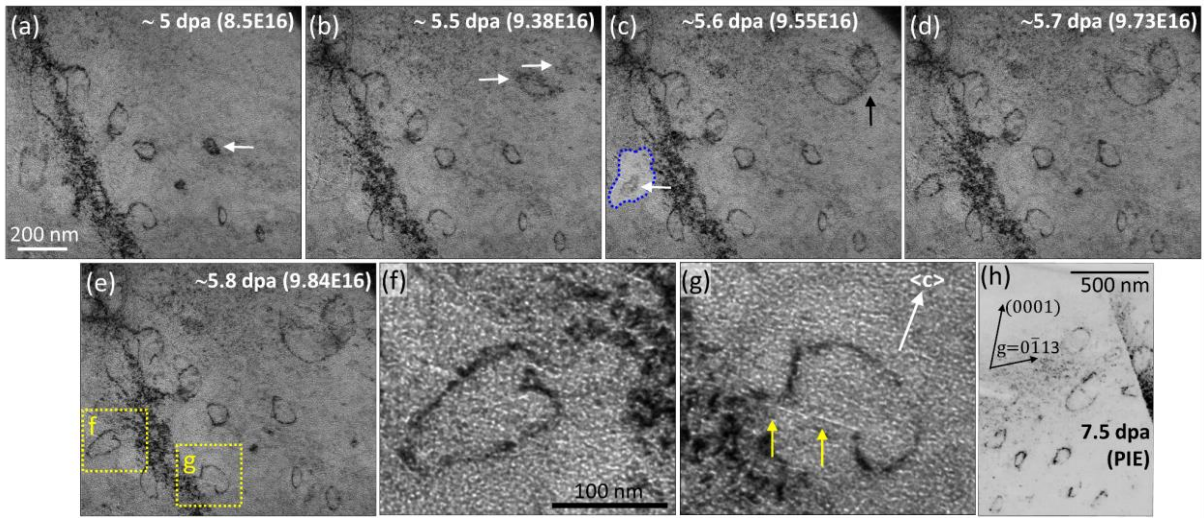

**Fig. 6.** (a-e) Snapshots from the *in situ* recorded BF TEM imaging video during an irradiation between 5.0 dpa and 5.8 dpa. White arrows (a-c) pinpoint the formation of new dislocation loops, the black arrow indicates a dislocation dipole. The shaded area in (c) highlights the growth of a former loop. Magnified images in (f-g) show the pinning of loops by bubbles and platelets, respectively. (h) Inverted DF image of dislocation loops.

Materials that show radiation swelling should contain, in theory, both cavities and dislocations that may act as sinks for irradiation-induced vacancies and interstitials, respectively [63]. It is possible that while the generated vacancies (thermally and/or radiation-induced) are occupied by the implanted He atoms, leading to bubbles and platelets, the irradiation-induced interstitial atoms continuously form new interstitial dislocation loops under irradiation, which are mobile and can grow at 500 °C.

## He⁺ irradiation at 600 °C

He$^+$ irradiation at 600 °C

At 600 °C, He platelets were first observed at the lower damage dose of 2.0 dpa, as compared to lower temperature irradiations where He platelets formed at higher doses (i.e., 5.0 dpa at 500 °C, and 7.5 dpa at 450 °C) (**Fig. 7b**). At 3.0 dpa, limited GB tearing started to occur (**Fig. 7c**). Spherical bubbles are visible in under- and overfocus BF images throughout the damage range. Similar to the findings at lower temperatures, He platelets are observed to align parallel to the basal planes (**Figs. 7e-7f**).

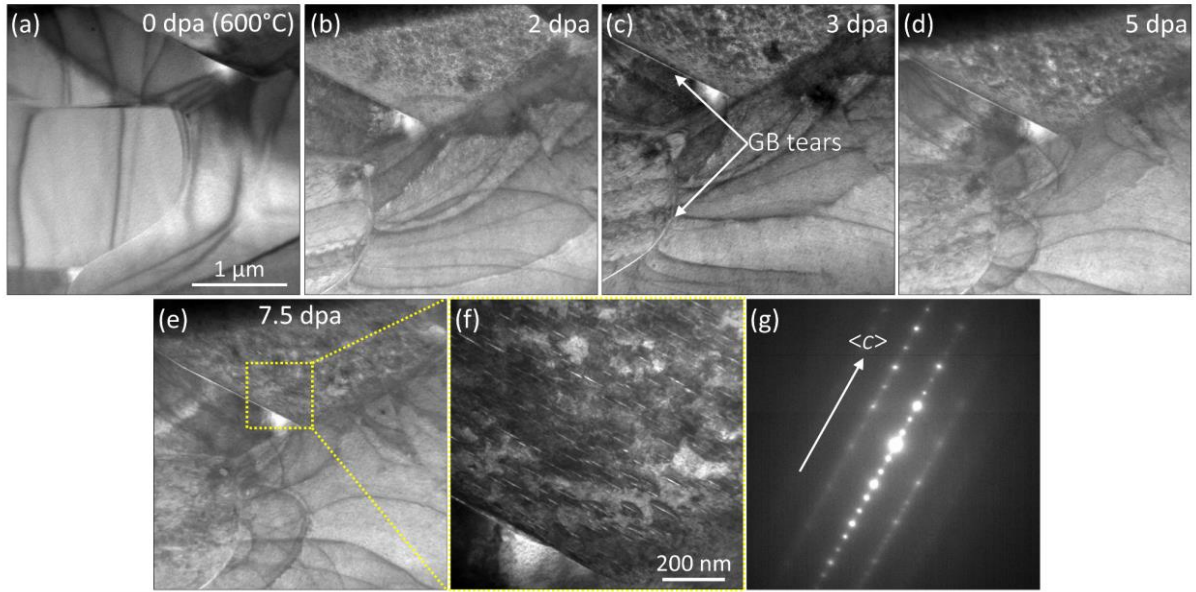

**Fig. 7.** BF TEM images at 600 °C before irradiation (a) , and after 2.0 (b), 3.0 (c), 5.0 (d) and 7.5 dpa (e). (f) The magnified image of the area that is framed in (e) shows He platelets formed parallel to the MAX phase basal planes; (g) corresponding SAED pattern of the same area.

## He⁺ irradiation at 700 °C

At 700 °C, He platelets formed at the lowest dose of 0.5 dpa (**Fig. 8a**), generating visible strain contrast, which became stronger with increasing platelet size (**Figs. 8b-8d**). Tearing of GBs was also observed, similar to what was observed at irradiation temperatures ≥ 450 °C. Before irradiation, dislocation loops in basal planes were visible (marked with arrows in **Fig. 8e**) in MAX phase grains at 700 °C. Such loops have been associated with the initiation of He and/or H platelet formation by attracting both vacancies and implanted He and/or H atoms, as described in literature for α-SiC [51]. The same area accommodated He platelets after irradiation to doses ≥ 0.5 dpa (**Figs. 8e-8h**). The diameters of the platelets marked with yellow arrows in **Figs. 8f-8h** have been measured in the 0.5-7.5 dpa range and are discussed further below.

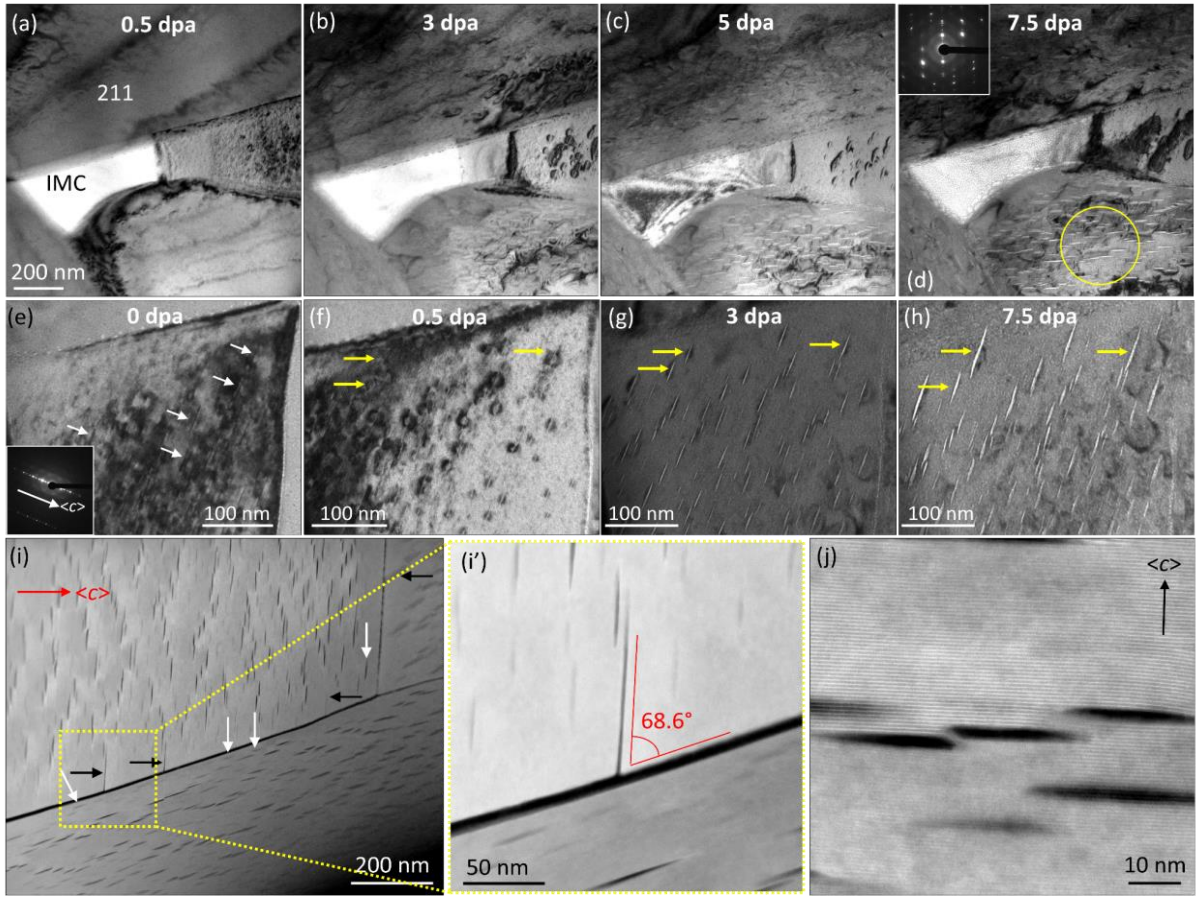

**Fig. 8.** *In situ* BF TEM images at 700 °C after (a) 0.5, (b) 3.0, (c) 5.0, and (d) 7.5 dpa. Another area at 700 °C at (e) 0, (f) 0.5, (g) 3.0, and (h) 7.5 dpa. At a damage dose of 7.5 dpa, HAADF-STEM images acquired during PIE show (i) a region with GBs, (i′) a magnified image of the area in the yellow frame in (i), and (j) a high-resolution image of He platelets.

During PIE, this thin foil was examined by means of high-resolution HAADF-STEM. With respect to the He platelets, no significant defect-denuded zones were observed, and the platelets were distributed uniformly within the grains, along the basal planes, and sometimes terminated on GBs with high angles to the basal plane (marked by black arrows in **Fig. 8i**). The magnified image of **Fig. 8i′** shows He platelets at an angle of ~68.6° to the GB (top grain) as well as platelets parallel to the GB (bottom grain). For GBs at low angles with the basal plane, either no significant He-platelet-denuded zone existed, or its width was limited in the 20–50 nm range. Moreover, the He platelets appeared very distorted at higher temperatures and in thinner regions of the foil. In **Fig. 8j**, the He platelets that lie parallel to the basal planes are visible and the nanolamination is preserved in the matrix around these defects. Increasing the platelet density did not result in platelet coalescence; therefore, He platelets were often observed on successive basal planes and at very close proximity to each other (**Fig. 8j**).

**Figures 9a** and **9b** show a HAADF-STEM image from a He-platelet-containing region and a magnified section revealing the irradiated lattice structure with the corresponding FFT image, respectively. At the $[2\bar{1}\bar{1}0]$ zone axis, as obtained from the FFT image, the nanolaminated structure is visibly present. The $\sim Z^2$ ($Z$ = atomic number) contrast of HAADF-STEM imaging indicates brighter (Al+Sn) layers and slightly darker (Zr+Ti) layers. The composition of the lattice and its individual layers might deviate locally from the $(Zr_{0.5},Ti_{0.5})_2(Al_{0.5},Sn_{0.5})C$ nominal bulk composition, either due to local compositional inhomogeneities that occur during sintering or due to the loss/segregation of certain elements resulting from irradiation and/or annealing. In **Figs. 9a-9b**, the bright $A$-layers suggest higher Sn and lower Al contents, since the mean atomic numbers of the $M$ and $A$ layers are comparable ($Z_{Zr+Ti}$=62, $Z_{Al+Sn}$=63), implying a small contrast difference. A prominent finding was observed in the EDX elemental mappings of a He-platelet-containing area (**Fig. 9c**). While the loss of elemental EDX signal in the Zr, Ti and Sn maps indicates the location of He platelets, the almost uniform, and locally slightly-enhanced, distribution of Al in the respective map indicates the presence of Al inside the He platelets after irradiation at 700 °C (**Fig. 9c**). This might suggest that Al interstitials (or Al adatoms) are present in the platelet volume. This hypothesis, however, needs confirmation by additional experiments and techniques, such as electron energy loss spectrometry (EELS). According to the annealing model proposed for $Ti_3SiC_2$ by Zhang $et$ $al.$ [6] and supported by various experimental studies, He will be accommodated in the Si layer at low He doses (200 keV, < $1\times10^{17}$ ions·cm$^{-2}$ [12]), whereas He clusters will burst due to an increased internal pressure above $1\times10^{18}$ ions·cm$^{-2}$ (50 keV [5]) and at < 1000 °C (decomposition of the MAX phase may occur above 1000 °C), followed by He release and Si-interstitial filling of the empty $A$-layers. Although the first-principles calculations aimed specifically at understanding the behaviour of $Ti_3SiC_2$, similar trends might be expected at high temperatures in other MAX phases, such as the $ZTAS$ studied in this work. However, such behaviour probably relates to cases where the laminated MAX phase crystal structure is preserved. A disrupted nanolamination due to platelet formation is responsible for the creation of free surfaces that are likely to be populated by Al interstitials, a fact that is expected to minimize the platelet surface energy and reduce local stresses caused by lattice interstitials. Helium desorption is also possible, considering the high mobility of He above 500 °C [20], the defective $A$-layers that may act as channels for fast elemental diffusion, and the thin foil sample geometry.

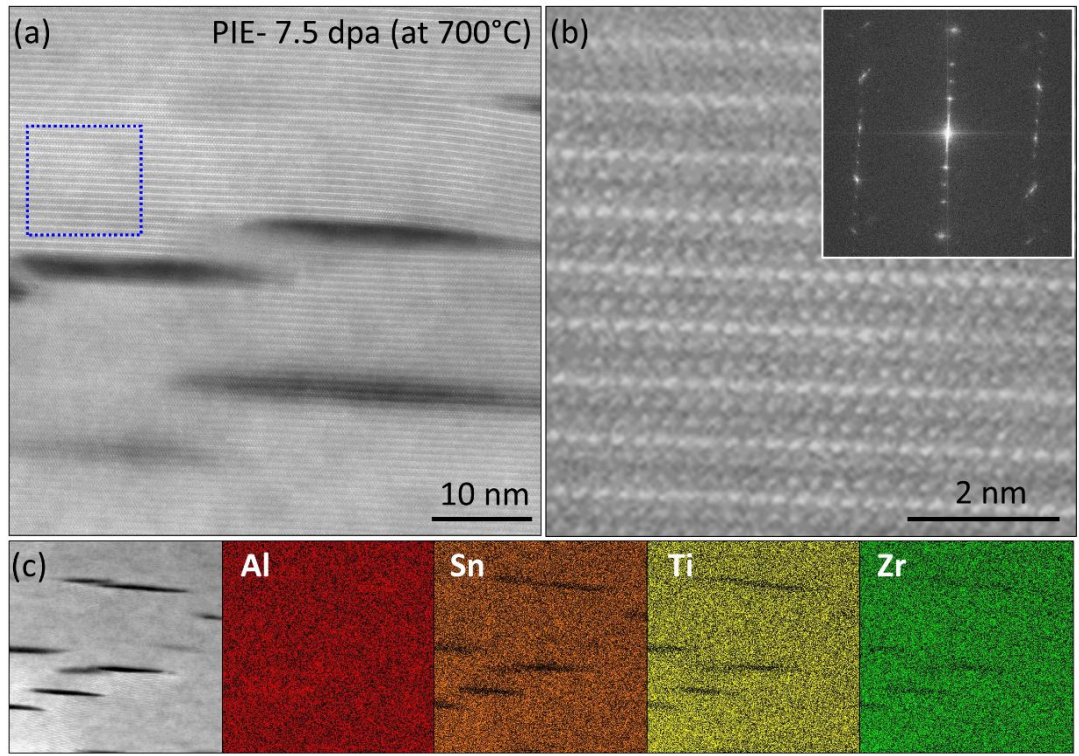

**Fig. 9.** (a) HAADF-STEM image collected during the PIE of a foil irradiated to 7.5 dpa at 700 °C. (b) Magnified image of the area framed in (a) with the FFT indicating a $[2\overline{1}\overline{1}0]$ zone axis. (c) Area with He platelets (7.5 dpa, 700 °C) with Al, Sn, Ti, and Zr STEM-EDX elemental maps.

## He⁺ irradiation at 800 °C

Similar to the observations at 700 °C, He platelets and nanosized spherical bubbles were observed at the lowest dose step of 0.5 dpa. **Fig. 10** shows the observed microstructural changes. Early He platelets appearing at 0.5 dpa showed the characteristic diffraction contrast resulting from the established strain field in their immediate vicinity (**Fig. 10a**), while at 1.0 dpa, uniformly distributed He platelets can already be observed (**Fig. 10b**). At 7.5 dpa, no He platelet- or spherical bubble-denuded zones were observed next to GBs (**Fig. 10c**). **Fig. 10d** presents a relatively thicker foil region with no traces of GB tearing. Higher magnification BF TEM images at ±2 µm defocus levels are shown in **Figs. 10e-10f**. The size of the spherical bubbles did not vary much when compared to lower irradiation temperatures. Similar to He platelets, He bubbles tend to align parallel to basal planes, as illustrated in the magnified image of **Fig. 10f'**. The linear alignment of spherical He bubbles might potentially be considered an initial (i.e., lower Gibbs free energy) stage in the formation of He platelets. However, the inverse process (disintegration of He platelets to bubbles) cannot be ruled out, as it has been observed in α-SiC, whereupon He platelets transformed to bubble-filed discs when α-SiC irradiated with He at RT was annealed in the 847-1247 °C range [53]. This transformation was the result of the disintegration of He platelets into small spherical bubbles, linearly arranged, and occupying the volume of the initial He platelet, thus maintaining the disc-like shape of the original platelet.

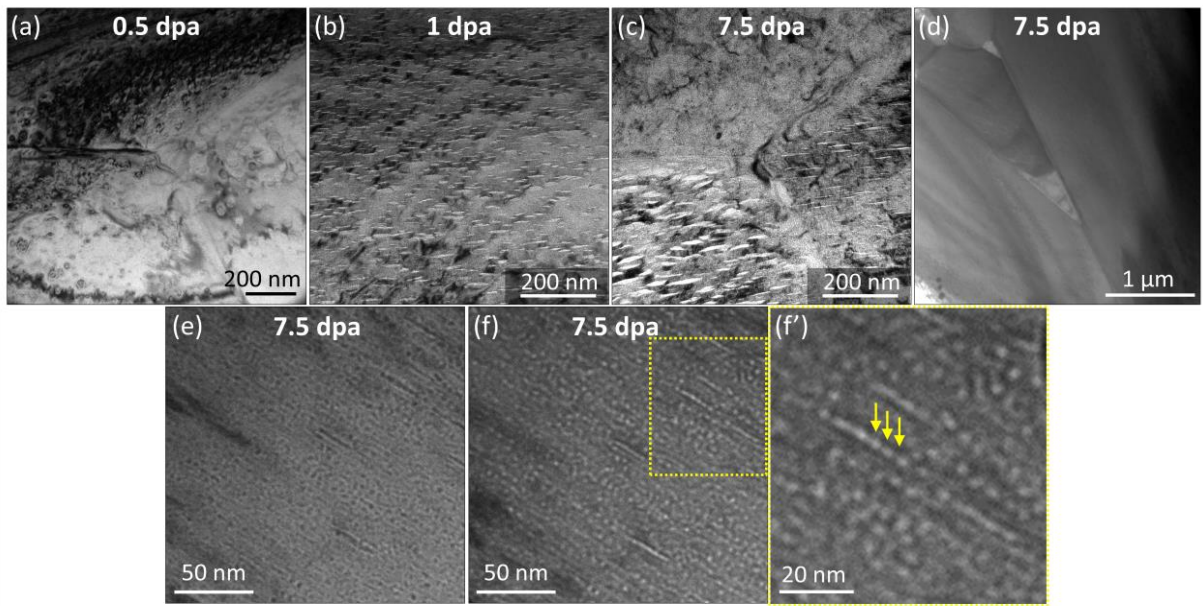

**Fig. 10.** BF TEM images at 800 °C after (a) 0.5, (b) 1.0 and (c–f) 7.5 dpa. At 7.5 dpa, $\pm2$ μm (e) overfocus and (f) underfocus images showing He bubbles and He platelets. Arrows in the magnified image (f') pinpoint linearly aligned spherical bubbles.

### Response of the $(Zr_{0.5},Ti_{0.5})_2(Al_{0.5},Sn_{0.5})C$ MAX phase solid solution to He$^+$ irradiation

Selected area electron diffraction patterns collected during the PIE of the *ZTAS* MAX phase after irradiation to 7.5 dpa at different irradiation temperatures are compared in **Fig. 11**. The SAED patterns presented in **Fig. 11** were collected from grains with better orientation (i.e., closer to low-order zone axes) and less FIB damage. No amorphisation was observed at the investigated temperatures and dose levels up to 7.5 dpa. Increasing the irradiation temperature to $\geq 450$ °C resulted in slight changes in the SAED patterns. The diffraction spots marked by the white arrows in **Figs. 11c** and **11d** started losing intensity, while the $(01\bar{1}l)$ ($l = 3n$) and $(000l)$ ($l = 6n$) spots, some of which are marked by red arrows in **Fig. 11d**, remained intense. The loss of intensity in select diffraction spots might be related to an increase in lattice symmetry towards an *fcc (M,A)C* structure due to chemical disordering in both *M* and *A* layers; such disordering has been commonly observed in MAX phase compounds irradiated to high dpa levels [5,9,21,31]. The SAED patterns in **Figs. 11c-11d** are comparable to those of *γ-Cr$_2$AlC* and *γ-V$_2$AlC*, a transition phase appearing before the formation of the *fcc-(M,A)C* phase at increased irradiation doses [64]. In this *γ-M$_2$AC* phase, C atoms are partly moved to the octahedral sites of the uniformly distributed *M/A* cation sites, resulting in the formation of antisite defects [64]. The SAED patterns collected from the foil irradiated at 800 °C (**Fig. 11**) showed no reduction in spot intensities and 'normal' (i.e., without signs of radiation damage) MAX phase SAED patterns were recorded.

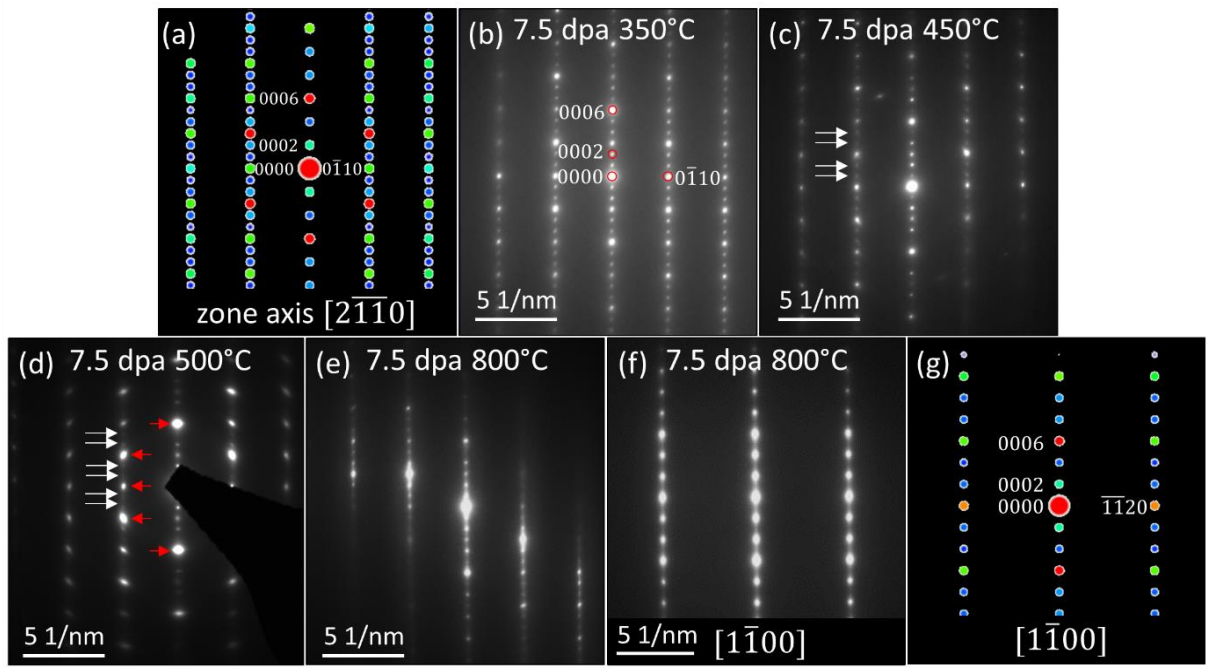

**Fig. 11.** (a) SAED patterns simulated for the *ZTAS* MAX phase at the $[2\overline{1}\overline{1}0]$ zone axis, alongside those recorded at the same zone axis during the PIE of *ZTAS* foils irradiated at (b) 350 °C, (c) 450 °C, (d) 500 °C, and (e) 800 °C up to 7.5 dpa. (f) SAED pattern after 7.5 dpa at 800 °C at the $[1\overline{1}00]$ zone axis, and (g) simulated SAED pattern at the same zone axis. White arrows show weakened diffraction spots, while red arrows indicate strengthened diffraction spots due to He⁺ irradiation.

The overall radiation-induced microstructural defects are presented as function of irradiation temperature and dose in **Fig. 12a**. The data points were plotted with a small offset to avoid overlapping. **Figures 12b-12c** present the temperature vs. dose maps for the presence of He strings/platelets and GB tears, respectively. The defects in these two plots were found to be critical for thin foil integrity; therefore, they were plotted separately to aid appreciation of the observed trends. Helium strings were not observed at 350 °C, but they appeared at intermediate temperatures (450-500 °C) and intermediate doses (2.0-5.0 dpa). Although He bubble strings are assumed to exist at 600, 700 and 800 °C as well, it was impossible to unambiguously identify them in the collected TEM images, except in the case of the foils irradiated to 1.0 and 7.5 dpa at 800 °C. Helium platelets were observed at high doses and intermediate temperatures, appearing after 7.5 dpa at 450 °C and 5.0 dpa at 500 °C. Their presence was also confirmed at lower doses and at temperatures ≥ 600 °C. At 800 °C, the lowest achieved dose (0.5 dpa) resulted in platelet formation. Grain boundary tearing was observed at early doses (3.0 dpa) at 500 and 600 °C, but appeared only at higher doses at 450 °C or ≥ 700 °C, as summarized in **Fig. 12c**. It should be noted that some of the observed features might have passed undetected at other doses or temperatures. This could be attributed to thickness variations of each individual thin foil and also to differences between the orientations of different grains viewed in TEM.

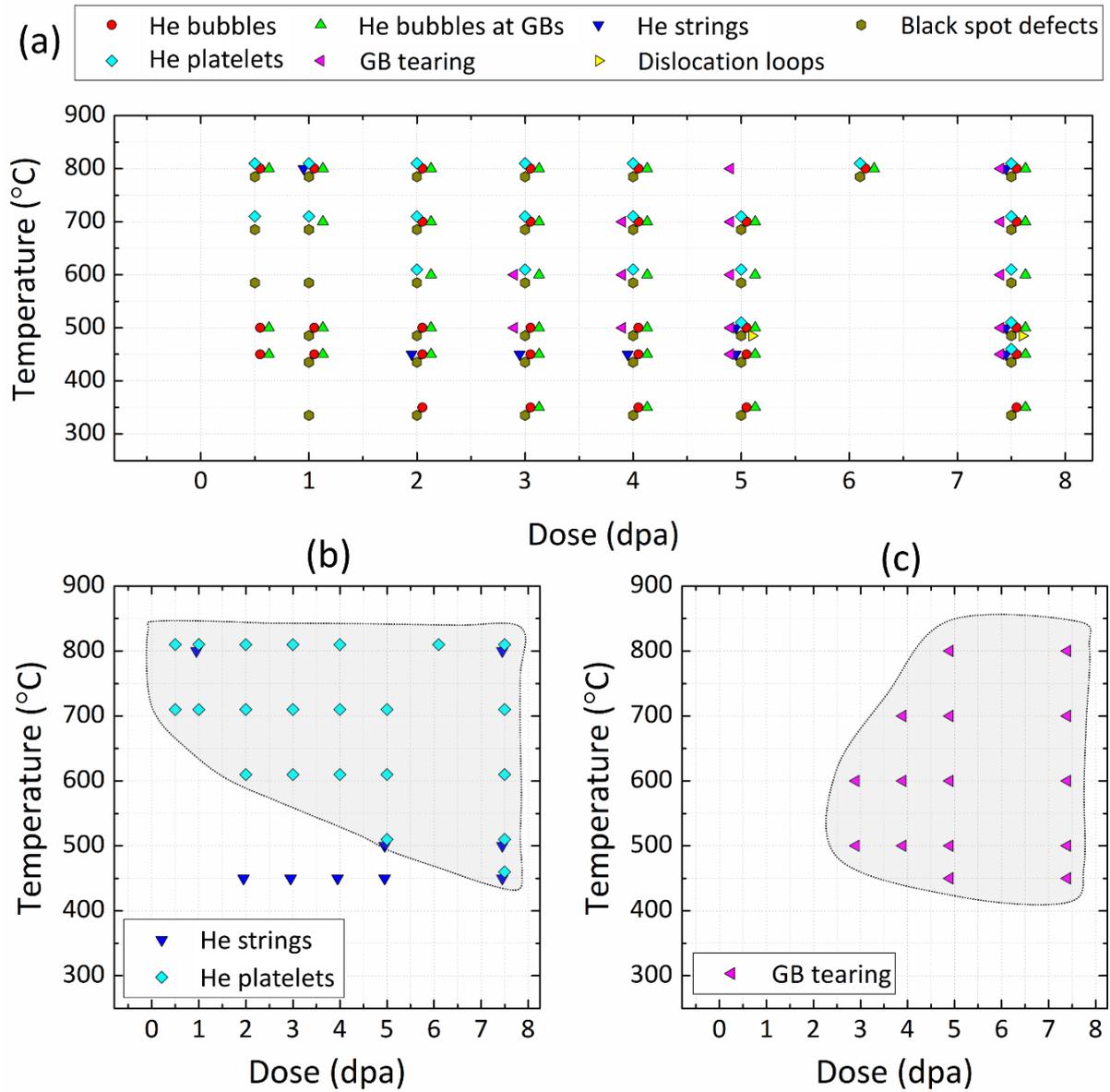

**Fig. 12.** (a) Overall microstructural defects as function of irradiation dose and temperature. Individual plots for (b) He strings/platelets, and (c) GB tearing. To avoid the overlap of data points, offsets along the *x* and *y* axes have been used in (a,b).

The He bubble diameter averaged in the 2.0-2.5 nm range and hardly varied with temperature, thickness of the investigated area (55-105 nm) and/or with further increase in the dose (**Fig. 13a**). First-principles calculations on $Ti_3AlC_2$ by Xu *et al.* [13] showed that He atoms trapped in a C-vacancy do not increase the likelihood of generating secondary C-vacancies. As a result, He atoms trapped in C-vacancies would prefer to grow into the other layers instead of growing along the C-layer. It was suggested that this behaviour will lead to the equiaxed growth of spherical He clusters [13]. Adjacent spherical bubbles can then migrate or combine to form string-like defects. The growth of spherical He bubbles was expected to be slow and limited, due to the high energy barrier associated with bubble growth along the <*c*> axis [13], which might also explain the stable bubble size observed. It was also calculated that compared to Al-vacancies, C-vacancies strain the lattice more when filled with two or more He atoms. Contrary to C-vacancies, He atoms residing in Al-vacancies were found to lower the energy required for the formation of secondary Al-vacancies, thus trapping more He atoms and allowing the 2D growth (i.e., platelet-like) of He clusters within the *A*-layers [13].

Comparing the average He platelet diameters in the foils irradiated at 450 °C and 700 °C, which had similar thicknesses (i.e., 94 and 108 nm, respectively), revealed an increase in diameter at the higher irradiation temperature at equivalent dose levels (**Fig. 13b**). However, similar foil thicknesses would be necessary for a proper comparison of He platelet diameters at different irradiation temperatures. It is thought that a He platelet growing within a very thin foil has only a limited volume to grow before reaching a free surface and losing the trapped He into the vacuum and thus releasing the pressure required for further platelet growth. It is therefore possible that only those He platelets which remain small can survive in thin areas, but in thicker areas they have greater space to grow in the in situ irradiation of thin foils reported here. The largest average platelet diameter measured was in the foil irradiated at 600°C up to 7.5 dpa, which happened to be the thickest examined area (185 nm). A relatively stable average platelet diameter of 25-35 nm was measured > 500 °C (except from the data point at 600 °C, 7.5 dpa). In SiC, a similar behaviour was observed upon He irradiation at RT, where He platelets with a narrow size distribution were found lying on (0001) basal planes [53]. Upon annealing up to ~1000 °C, the platelet size and thickness remained almost constant, while the number density increased. This was explained by the low mobility of dislocation dipoles generated at the rims of the He platelets due to displaced atomic planes above and below the platelets, which limited the platelet growth [53]. Growth kinetics of He platelets in Si, under H supply, were studied in the work of Vallet *et al.* [65]. They measured the sizes of individual He platelets and found that the platelet growth follows the Johnson-Mehl-Avrami-Kolmogorov (JMAK) model. Similar work on the MAX phases under irradiation has never been reported. Inspired by the work of Vallet *et al.* [65], the diameters of three individual platelets (marked with yellow arrows in **Figs. 8f-8h**) were measured at each dpa level at 700 °C, indicating linear He platelet growth (**Fig. 13c**). The 700 °C irradiation experiment was the only experiment where these measurements could be performed from the collected TEM images, allowing the diameter measurement of the same He platelets throughout the full dpa range. The observation of a quasi-stable average He platelet size at each temperature (**Fig. 13b**), when individual platelets showed linear growth with dose (**Fig. 13c**), could be attributed to the continuous nucleation of He platelets during the ion irradiation experiments. Nucleation of new (initially small) He platelets is considered more favourable than the growth of existing ones, as the latter might lead to the delamination of basal planes, causing severe material damage.

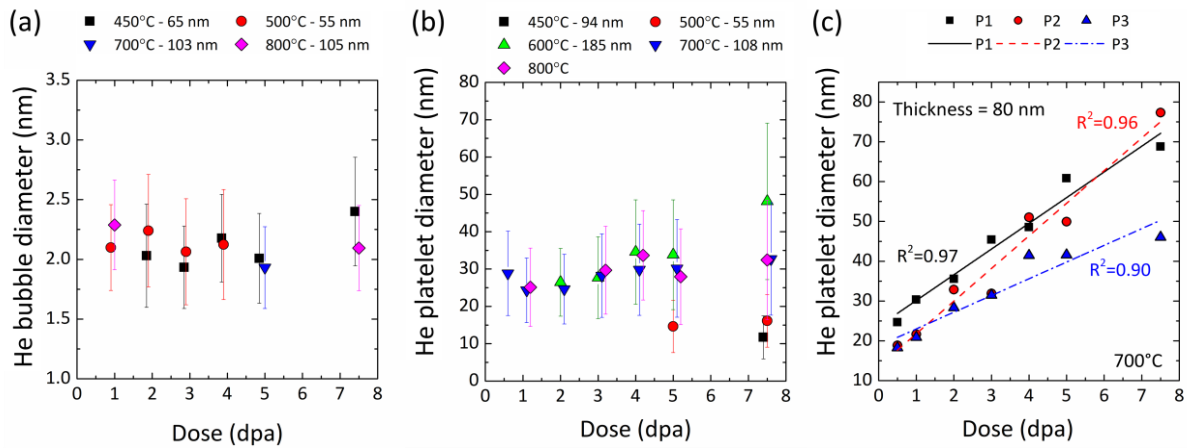

**Fig. 13.** Average diameter of (a) spherical He bubbles and (b) He platelets as function of dose and temperature. (c) Diameter evolution of three individual He platelets (marked in **Figs. 8f-8h**) at 700 °C at different dose levels. To avoid data point overlap, offsets were used along the *x*-axis in (a,b). Foil thicknesses are indicated in the plot legends.

With a simple approach, the stable size of spherical bubbles shown in **Fig. 13a** might be attributed to the fact that He contributes mainly to He platelet growth. However, understanding the dose effect on the growth of spherical He bubbles or He platelets (here named 'cavities') would have been easier, if only one of these two mechanisms was operative. The effect of the amount of implanted He (i.e., the dose) on the He cavity size should ideally be investigated by considering the density and size of both types of cavities in the same area, since the increase in the size/density of one cavity type would imply less available He (at the same dose) for the growth of the other cavity type. Thus the information in **Fig. 13** should only be considered as a rough guide for the baseline cavity sizes obtained in this study; similar defects are expected in ion irradiations of bulk *ZTAS* MAX phase ceramics under similar conditions, without the thin foil surfaces acting as defect sinks.

For specific MAX phases, saturation of the ion irradiation-induced damage has been detected by a plateau in their hardness values above a certain irradiation dose [23,66]. In this work, measuring hardness was impossible; moreover, no systematic defect density determination has been performed. However, He platelet growth was invariably observed as function of dose (**Fig. 13d**) for the investigated ion irradiation conditions. Therefore, saturation of radiation-induced damage cannot be claimed for the *ZTAS* double solid solution MAX phase investigated in this work.

Regarding the effect of the IMC phase on the overall response of the *ZTAS* MAX phase solid solution to irradiation, no significant effects were noticed within the investigated window of irradiation temperatures and doses. GB tears did not uniquely occur between IMC and MAX phase grains, so as to unequivocally consider the IMC detrimental on the radiation tolerance of the *ZTAS* ceramic. However, this might be associated with the limited IMC fraction in the foils and/or the irradiation temperature and dose ranges studied in this work. The IMC also contained He bubbles and He platelets at various doses and temperatures, but the response of IMCs to He irradiation has not been studied systematically in this work. In the supplementary information (**Fig. S6**), He bubbles and platelets observed in the IMC are shown. The formation of He bubbles and platelets is expected to induce a certain degree of radiation swelling to the IMC.

In the 550 °C He$^+$ irradiation experiment, where the fluence control was lost due to ion beam instability, the thin foil was damaged to an unknown level, which was likely much higher than 7.5 dpa. A number of BF TEM images obtained during this experiment are presented in the supplementary information (**Fig. S7**) as examples of severe irradiation damage. Combined He platelets leading to delaminations and microcracks across GBs and along the grains were observed. Complete detachment (GB tearing) of an IMC grain from the MAX matrix is visible in **Fig. S7**. Considering the potential differences in thermal expansion coefficients and irradiation-induced swelling of IMCs and MAX phases that might lead to material cracking in service (i.e., in the reactor), it is safer to completely eliminate the IMC phases (if possible) by optimizing the synthesis parameters to achieve phase-pure MAX phase ceramics. It should also be noted that the amount of He implanted during these experiments, so as to reach the targeted dose level of 7.5 dpa, does not represent the actual in-service He/dpa ratio of a nuclear fuel cladding, which will be substantially lower. However, the added value of such ion irradiation experiments lies in the accelerated assessment of the radiation swelling behaviour of innovative candidate nuclear materials, such as the *ZTAS* MAX phase studied in this work.

The reference foils that were included in the 350 and 500 °C irradiation experiments to solely assess the thermal stress effects showed no visible damage. This indicated that the anisotropic thermal expansion/contraction did not lead to GB tearing or microcracking. Therefore, the observed GB tears in the irradiated foils are thought to have originated from the accumulation of He bubbles at the GBs of the irradiated *ZTAS* MAX phase grains. Although observed in most thin foils, these tears did not propagate further as transgranular cracks during the irradiation or upon cooling, but remained rather confined as local damage at the GBs.

## Conclusions

Quasi phase-pure $(Zr_{0.5},Ti_{0.5})_2(Al_{0.5},Sn_{0.5})C$ MAX phase thin foils were *in situ* irradiated in a TEM at 350, 450, 500, 600, 700 and 800 °C, using 6 keV He$^+$ up to 7.5 dpa ($1.3\times10^{17}$ ions·cm$^{-2}$) with an ion flux of $8.8\times10^{13}$ ions·cm$^{-2}$·s$^{-1}$.

- Spherical He bubbles, stable in size (approximately 2.0-2.5 nm) were observed at all doses ($\geq$ 0.5 dpa) and temperatures with a uniform distribution within the grains and at the GBs. A high accumulation of He bubbles resulted in GB tearing $\geq 450\,°C$, but no crack propagation was observed.

- Helium string-like bubbles and platelets were observed starting at $450\,°C$ at the highest dose of 7.5 dpa. At higher temperatures, they became visible at lower damage dose levels. At $700\,°C$, linear He platelet growth (i.e., increase in diameter) was observed with increasing damage dose. Despite the observed scatter in the measured average He platelet diameter data, the irradiation temperature seems to be the dominant factor in determining the He platelet diameter rather than the dose (accumulated at a constant temperature), as similarly thick foils were characterised by increased platelet diameter as the temperature increased at a specific dose.
- At $500\,°C$, mobile dislocation loops were recorded as evidence of irradiation-induced defect mobility. Higher defect mobility can be interpreted as potentially higher defect recombination and migration to sinks, indicating promising radiation tolerance.
- The slight decrease in the intensity of some diffraction spots in the SAED patterns after irradiation at 450 and $500\,°C$, which indicates transformation to a $\gamma$-$M_2AC$ phase, was recovered at higher temperatures: HRSTEM images showed retention of chemical ordering and nanolamination after irradiation to 7.5 dpa at $700\,°C$.
- The evolution of the observed radiation-induced defects was mapped as a function of irradiation temperature and damage dose.
- No amorphisation was observed. Uniformly distributed He bubbles/platelets of stable size and mobile dislocations are believed to act as defect sinks, suggesting that He is stored in the $(Zr_{0.5},Ti_{0.5})_2(Al_{0.5},Sn_{0.5})C$ MAX phase ceramic without lattice disruptions and/or severe material degradation.

Although *in situ* ion irradiation gives valuable information on the dynamic evolution of radiation-induced defect microstructures, the surfaces of the thin foils act as defect sinks that might misrepresent the actual irradiation response of the corresponding bulk materials. More irradiation studies, including irradiations of bulk materials, are necessary to understand the radiation-induced defect evolution in MAX phase ceramics intended for nuclear applications.

## Acknowledgements

B.T. acknowledges J.W. Seo for the TEM training and guidance, as well as the financial support of the SCK CEN Academy for Nuclear Science and Technology. This research has been funded by the Euratom research and training programme 2014-2018 under Grant Agreement No. 740415 (H2020 IL TROVATORE). The performed research falls within the framework of the EERA (European Energy Research Alliance) Joint Programme on Nuclear Materials (JPNM). The authors gratefully acknowledge the Hercules Foundation for project AKUL/1319 (CombiS(T)EM), UK's Engineering and Physical Sciences Research Council (EPSRC) for funding the construction of the MIAMI facility under Grant EP/M028283/1, as well as the Swedish Research Council for funding under Grant no. 2016-04412, the Swedish Foundation for Strategic Research (SSF) through the Research Infrastructure Fellow program no. RIF 14-0074.

The authors finally acknowledge support from the Swedish Government Strategic Research Area in Materials Science on Functional Materials at Linköping University (Faculty Grant SFO-Mat-LiU No 2009 00971) and the Knut and Alice Wallenberg's Foundation for support of the electron microscopy laboratory in Linköping.